\documentclass[a4paper,fleqn,usenatbib]{mnras}

\usepackage[T1]{fontenc}
\usepackage{ae,aecompl}

\usepackage{natbib}
\usepackage{graphicx}
\usepackage{enumitem}
\usepackage{calc}
\usepackage{lipsum,adjustbox}
\usepackage{subcaption}
\usepackage{tikz}
\usepackage{amsmath}	% Advanced maths commands
\usepackage{amssymb}
\usepackage{wasysym}
\usepackage{txfonts}
\usepackage{array}
\usepackage{threeparttable}
\usepackage{float}
\usepackage{dcolumn}
\usepackage{epstopdf}

\title[Improved Asteroseismic ConvNets for K2/TESS]{Deep Learning Classification in Asteroseismology Using an Improved Neural Network: Results on 15000 \textit{Kepler} Red Giants and Applications to K2 and TESS Data}

\author[Hon et al.]{
	Marc Hon,$^{1}$\thanks{E-mail: mtyh555@uowmail.edu.au}
	Dennis Stello,$^{1,2,3}$ 
	and Jie Yu$^{2}$
	\\
	$^{1}$School of Physics, The University of New South Wales, Sydney NSW 2052, Australia\\
	$^{2}$Sydney Institute for Astronomy (SIfA), School of Physics, University of Sydney, NSW 2006, Australia\\
	$^{3}$Stellar Astrophysics Centre, Department of Physics and Astronomy, Aarhus University, Ny Munkegade 120, DK-8000 Aarhus C, Denmark\\
}

\date{Accepted XXX. Received YYY; in original form ZZZ}

\pubyear{2017}

\begin{document}
	\label{firstpage}
	\pagerange{\pageref{firstpage}--\pageref{lastpage}}
	\maketitle
	
	\begin{abstract}
		Deep learning in the form of 1D convolutional neural networks have previously been shown to be capable of efficiently classifying the evolutionary state of oscillating red giants into red giant branch stars and helium-core burning stars by recognizing visual features in their asteroseismic frequency spectra. We elaborate further on the deep learning method by developing an improved convolutional neural network classifier. To make our method useful for current and future space missions such as K2, TESS and PLATO, we train classifiers that are able to classify the evolutionary states of lower frequency resolution spectra expected from these missions. Additionally, we provide new classifications for 8633 \textit{Kepler} red giants, out of which 426 have previously not been classified using asteroseismology. This brings the total to 14983 \textit{Kepler} red giants classified with our new neural network. We also verify that our classifiers are remarkably robust to suboptimal data, including low signal-to-noise and incorrect training truth labels.

	\end{abstract}
	
	
	%%%%%%%%%%%%%%%%%%%%%%%%%%%%%%%%%%%%%%%%%%%%%%%%%%
	
	%%%%%%%%%%%%%%%%% BODY OF PAPER %%%%%%%%%%%%%%%%%%
	
	\section{Introduction}
	In constraining the stellar ages of red giants, it is important to determine their evolutionary state and distinguish between those that are currently burning hydrogen in a shell surrounding their inert helium core and those that have already commenced core-helium burning. The former are known as red giant branch (RGB) stars, while the latter are helium burning (HeB) or clump stars. Both types have overlapping populations in the colour-magnitude diagram, which makes it challenging to distinguish them using classic stellar surface properties such as luminosity and effective temperature. By observing the stochastically excited and damped, so-called solar-like, oscillation frequencies of red giants, asteroseismology effectively probes the interior stellar structure and has enabled classifications based on the observed dipole mode period spacings (e.g \citealt{Bedding_2011, Stello2013}), the asymptotic dipole mode period spacings (e.g \citealt{Bedding_2011,Mosser_2012, Mosser2014, Vrard2016}), and mixed mode frequency distributions \citep{Elsworth2016}.
	
	In our previous work \citep[][hereafter H17]{Hon_2017}, we introduced an asteroseismic deep learning method using 1D convolutional neural networks. The method classifies the oscillation spectra of red giants into RGB or HeB stars by recognizing the significant visual patterns shown in the frequency power spectra by the oscillation modes that are characteristic to the evolutionary state of the red giant. This automated method uses visual recognition to classify power spectra and is conceptually similar to computer vision applications such as facial recognition (e.g  \citealt{Garcia_2004}), image detection and classification (e.g. \citealt{Krizhevsky_2012, Simonyan_2014}), text categorization (e.g. \citealt{Zhang_2015}), and biomedical image analyses (e.g. \citealt{Ciresan_2013, Prasoon_2013}). 
	
	With the advent of space missions K2 \citep{Howell_2014}, TESS \citep{TESS}, and PLATO \citep{Rauer_2014}, the need for wide-scale red giant classification becomes ever more demanding, because the total amount of red giant data received surpasses the amount from the \textit{Kepler} mission \citep{Borucki2010} by orders of magnitude \citep{Campante_2016, Miglio_2017, Stello_2017}. Moreover, the shorter time series from these newer missions make `classical' asteroseismic methods for classification more challenging, because they do not have sufficient frequency resolution to resolve the complex frequency structure of the dipole modes. Hence, there exists a need to expand our method to missions like K2 and TESS. In doing so, we will be better prepared to analyse and interpret the incoming large volume of data efficiently and effectively.
	
	\begin{figure*}
		\centering
		\includegraphics[width=0.8\linewidth]{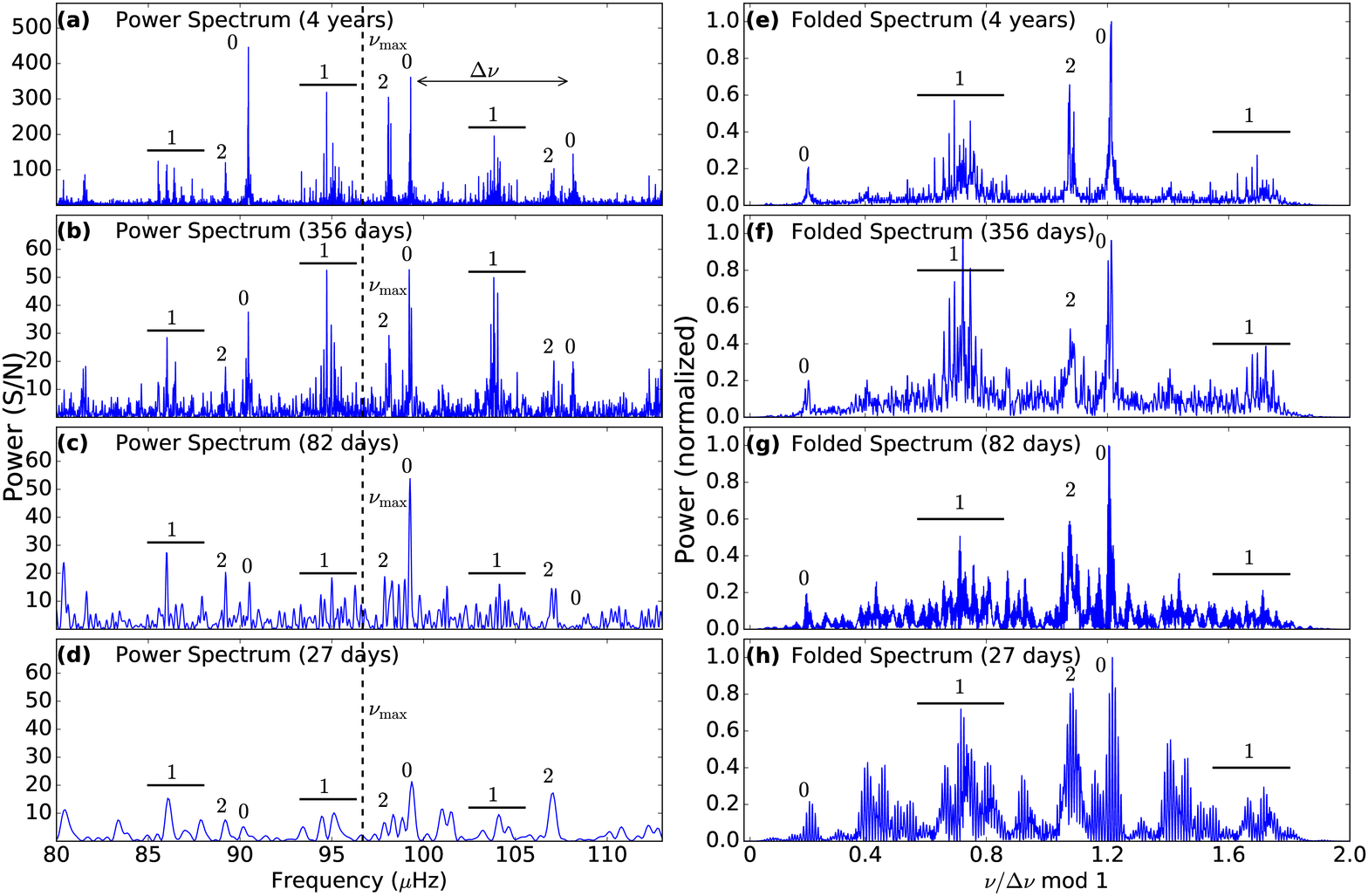}
		\caption{Noise normalised power spectra (left) and corresponding folded spectra (right) of the RGB star KIC 3641504 for varying lengths of observation time. The frequency at maximum power, $\nu_\mathrm{max}$, the overtone frequency separation, $\Delta\nu$, and the spherical degrees of the low degree acoustic modes are indicated. (a, e) The observation length of 4 years is representative of the full \textit{Kepler} mission, (b, f) the length of 1 year is typical for the TESS continuing viewing zones, (c, g) the length of 82 days is typical for a K2 campaign, (d, h) and 27 days is representative of the shortest TESS observations.}
		\label{DifferentTimes}
	\end{figure*}
	
	In this paper, we expand on the deep learning method developed by H17 and introduce an improved version of the convolutional neural network classifier. We train versions of this new classifier on data lengths specific to \textit{Kepler}, K2, and TESS. Using these trained classifiers, we provide new classifications for the \textit{Kepler} red giants and predict the evolutionary states of the red giants within the open cluster M67 using K2 data. Finally, we investigate each classifier's performance when training and predicting on suboptimal data. We refer the reader to H17 for a thorough introduction to the techniques, concepts, and typical naming conventions in deep learning aimed at a non-specialist audience.

	%4755
	\section{Data and Performance Testing Methods}\label{Data}
	Similar to H17, we use the \textit{folded spectrum} as an image representation for the data in this study (Figure \ref{DifferentTimes}). The folded spectrum is a 4$\Delta\nu$-wide power spectrum segment centered at $\nu_\mathrm{max}$ that is folded by a length of $\Delta\nu$, then appended with a copy of itself and weighted with a super-Gaussian to smoothly tamper down the edges (see H17 for details). The folded spectra are constructed from power spectra that are divided by their background noise, making them flat with a mean noise level of 1.0. To train and measure the performance of classifiers, we use a dataset comprised of 6015 \textit{Kepler} red giants with classifications based on automated period spacing measurements from \citet{Vrard2016} and an additional 335 stars from the classification from \citet{Mosser2014} to a total of 6350 unique stars, with a ratio of RGB to HeB stars of approximately 1:2. We use the same 5000 stars as H17 as the training set, with the remaining 1350 stars as the test set. We assign RGB stars with the binary class 0 and HeB stars with class 1. 
	
	\subsection{Shorter Time Series}\label{shorterseries}
	
	To simulate the time series for K2 and TESS observations, we split the existing \textit{Kepler} time series into multiple segments of shorter length. For K2, we used light curve segments of 82-day duration, while for TESS, we used 356-day, 82-day, and 27-day segments. Combined, the results from \textit{Kepler} and our K2 and TESS simulations form a representative set for what one can expect from PLATO, depending on the observational strategy the mission will adopt. The effect of a shorter observation time on the quality of the power spectrum is illustrated in Figure \ref{DifferentTimes}. With shorter observations, the frequency resolution of the data decreases, the signal-to-noise for oscillation peaks decreases, and the visibility of resolvable peaks is reduced. While splitting the time series into smaller segments produces data of lower quality, it also produces multiple time series originating from the same star. This can potentially improve the performance of the classifier because the number of stars within the training set can be artifically boosted by having multiple `independent' time series segments of a star within the dataset.
	These segments can be considered independent because they contain natural variations in noise and mode visibilities amongst themselves, such that their image representations are not too similar. We have verified that the classifiers do not produce the exact same prediction for multiple segments from the same star. Thus, the power spectra of multiple segments from the same star contain a sufficient level of variation such that they can be considered as different stars from another. The inclusion of these multiple segments in training allows the classifier to generalize better. We denote a dataset containing multiple segments from the same star as a \textit{degenerate} set. We obtain degenerate training sets containing 14110, 66376, and 193534 stars for time series lengths of 356 days, 82 days, and 27 days, respectively. 
	
	In contrast, we denote a dataset without multiple segments of the same star as a \textit{non-degenerate} set. In other words, all stars in a non-degenerate set have unique KIC identifiers within the set. We only use non-degenerate sets for the test set. Because the visual representation of the folded spectra can appear significantly altered by the changes in data quality across different lengths of time series, we train separate classifiers for each data length. Test sets of a particular length of time series will only be predicted on by a classifier that is trained on a degenerate training set of an identical length of time series. In this paper, we commonly use the term `data length' for a particular dataset. This term references the length of the time series from which the dataset is created, not the dimensionality of the dataset.
	
	\subsection{$\Delta\nu$ Precision}\label{DnuPrecision}
\begin{figure}
		\centering
		\includegraphics[width=\linewidth]{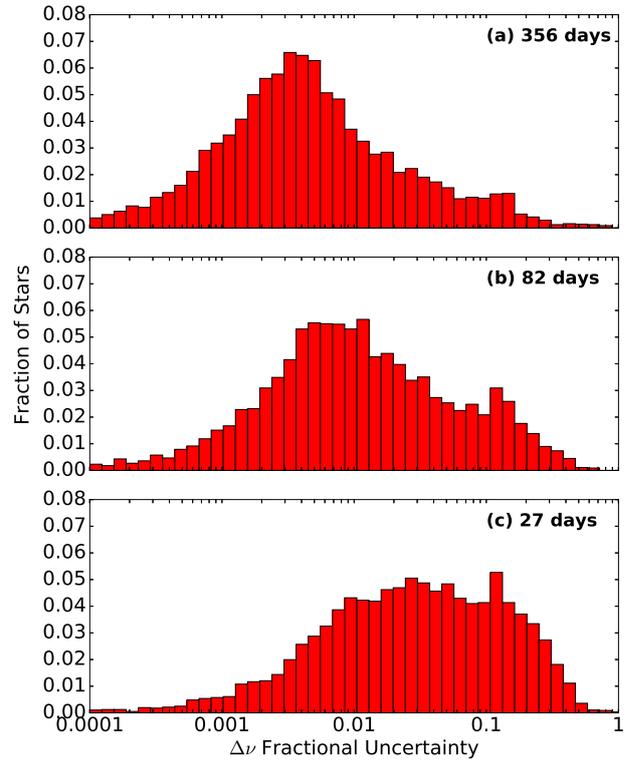}
		\caption{The distribution of SYD pipeline fractional uncertainties for $\Delta\nu$ for time series of shorter lengths, measured as their fractional deviation from corresponding 4-year $\Delta\nu$ values.}
		\label{DNUFractionalError}
	\end{figure}

	As in H17, we obtain the $\Delta\nu$ values from the SYD pipeline \citep{SYD} in order to generate the folded spectra. For the 4-year \textit{Kepler} data we adopt the $\Delta\nu$ results from \citet{Yu_2018}. However, with shorter time series, we expect the $\Delta\nu$ values to be less precise. Because we want our datasets to be representative of K2 and TESS data not just in terms of frequency resolution but also in terms of $\Delta\nu$ precision, we simulate $\Delta\nu$ measurements for shorter data lengths. First, we measure the fractional deviation of 6090 stars' 356-day, 82-day, and 27-day $\Delta\nu$ values from their 4-year values, which we show in Figure \ref{DNUFractionalError}. While this set of stars is only a subset of the total number of classified red giants ($\simeq15000$), they are approximately evenly distributed in evolutionary state population, and are only selected because they show a high level of continuity in each shorter time series segment.
	Next, to generate representative $\Delta\nu$ values for each 356-day, 82-day, and 27-day datasets, we let the $\Delta\nu$ value for each star be normally distributed about its 4-year $\Delta\nu$ value, with a standard deviation sampled from their corresponding uncertainty distributions in Figure \ref{DNUFractionalError}. In order to obtain representative $\Delta\nu$ values for each star, we sample from each of their normal distributions. Using this method, multiple segments of the same star in a degenerate set (as discussed in Section \ref{shorterseries}) will have $\Delta\nu$ values slightly differing from one another. We now use these representative $\Delta\nu$ values to create our training and test sets for shorter time series.

	\subsection{Testing the Classifier's Performance}\label{TestingPerformance}
	We test the performance of a classifier using two general methods. The first is by testing on the test set, which provides an unbiased measure of performance because the classifier does not train on the test set. Another method is to partition the training data into $k$ separate sets or folds, then train on $k-1$ (training) folds and measure the performance on the remaining (validation) fold. This method is known as \textit{k-fold cross validation}, and is useful because measuring the average metrics over $k$ independent sets can better account for the variance from predicting on a finite number of samples. We do not include the test set when performing k-fold cross validation because we also use cross-validation to tune the parameters of the network. Hence, including the test set would likely cause our classifier to have parameters that are tuned specifically to perform well on the test set. However, the test set will then no longer be an unbiased measure of generalization performance of the classifier. A summary of the metrics we use to describe the classifier perfomance is as follows:
	 
	 \begin{description}[]
		 	\item[Accuracy:] The number of correct predictions out of all predictions. 
		 	\item[Precision(P):] For a class, the ratio of correct predictions to \textit{all made predictions} towards that same class. Here it is the classifier's ability to not label a HeB star as an RGB star.
		 	\item[Recall(R):] For a class, the ratio of correct predictions to \textit{all stars} truly in that same class. Here it is the classifier's ability to find all HeB stars.
		 	\item[F1 Score:] The harmonic mean of precision and recall, defined by $2\frac{P\times R}{P + R}$, with 1 as a perfect score.
		 	\item[ROC AUC:] \textbf{R}eceiver \textbf{O}perating \textbf{C}haracteristic's \textbf{A}rea \textbf{U}nder \textbf{C}urve, which measures the classifier's average performance across all possible score thresholds. Has a value of 1 for a perfect classifier. 
		 	\item[Log Loss:] Negative logarithm of prediction scores i.e. the \textit{cross entropy}. Measures how well prediction scores are calibrated with an ideal value of 0 (see Equation \ref{LogLoss}).
		 	\item[Brier Score:] Mean squared error between predicted probability and ground truth (see Equation \ref{Brier}).
	 \end{description}
	
	\section{The New and Improved Classifier} 
	\begin{figure}
		\centering
		\includegraphics[width=1.0\linewidth]{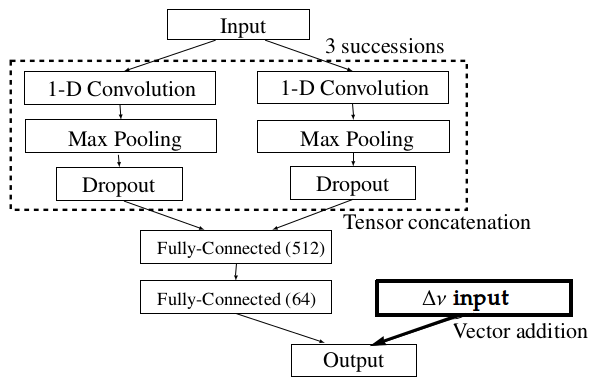}
		\caption{Schematic of the structure of the altered deep learning classifier, denoted as the \textit{new network}. The structure is similar to the classifier structure by H17, except here values of $\Delta \nu$ are added to the network (the bolded diagram branch). The numbers in brackets indicate the number of neurons in the layer.}
		\label{AdditionModel}
	\end{figure}
	In this Section, we introduce a new and improved classifier structure, and optimize it by determining the ideal probability threshold for separating RGB and HeB stars. 
	
	\subsection{New Classifier Structure}\label{ClasStructure}
	In our previous work, we developed a deep learning classifier, which we denote here as the \textit{old network}. Although the old network could predict the evolutionary phases of \textit{Kepler} red giants with a high degree of accuracy, it provided confident HeB predictions for stars with high $\Delta \nu$ ($ \apprge 9 \mu$Hz), a range where HeB stars should not exist (e.g \citealt[][their Figure 4b]{Stello2013}, and \citealt[][their Figure 1]{Mosser2014}). Its overconfidence for such predictions may also have indicated that it was overfitting the data during training.
	
	To address this limitation, we develop a variation of the old network that predicts mainly RGB stars for $\Delta \nu \apprge 9 \mu$Hz. We achieve this by providing $\Delta \nu$ as an additional input at the output layer of the classifier as illustrated in Figure \ref{AdditionModel}. This altered version is denoted as the \textit{new network}. The effect of explicitly providing the network, or classifier, with values of $\Delta \nu$ is to provide a prior on the expected distribution of RGB and HeB stars, such that the classifier recognizes ranges of $\Delta \nu$ where certain populations do not occur. 
	
	 We evaluate the performance of our new network using the log loss metric, which measures the dissimilarity between the prediction probability, $\hat{y}$, with the ground truth $y$, given by:
	\begin{equation} \label{LogLoss}
	E(\mathbf{y, \hat{y}}) = -\frac{1}{m} \sum_{i=1}^{m} \bigg[y_i\log\hat{y}_i + (1-y_i)\log(1-\hat{y}_i)\bigg],
	\end{equation}
	where $\hat{y}$ is a classification probability between 0 and 1, and $y$ is either 0 or 1. The lower the value of the log loss, the better the performance of the classifier in classifying stars. Reporting the log losses allows us to examine the performance of the classifier in terms of their probabilities. This is a more informative metric than the accuracy metric because the latter only evaluates the number of correct predictions without considering the degree of belief or probability for each prediction. For this evaluation, we bin the stars into three specific $\Delta \nu$ ranges according to the occurrence of particular red giant populations: mostly RGB ($\Delta \nu > 9\mu$Hz), RGB and the secondary clump ($5\mu$Hz $< \Delta \nu \le 9\mu$Hz), and RGB and the red clump ($\Delta \nu \le 5\mu$Hz). 
	\begin{figure}
		\centering
		\includegraphics[width=1.0\linewidth]{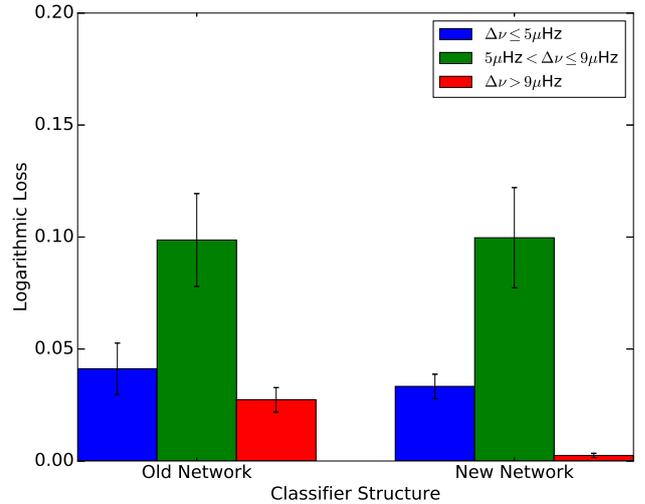}
		\caption{Log loss comparison between 4-year classifiers constructed using the old and new networks. The log loss is evaluated for stars within specific $\Delta\nu$ ranges by 10-fold cross validation, with the uncertainties as the standard errors.}
		\label{LogLossSingleTrio}
	\end{figure}
	 We show the log loss comparison between the old and new networks based on 4-year data in Figure \ref{LogLossSingleTrio}. The comparison of log loss values for models of shorter data lengths are very similar to that of the 4-year models and hence not shown. By comparing the log loss between the old and new networks, we see that the new network has a significantly smaller log loss for $\Delta\nu > 9\mu$Hz (red), which is due to the use of the $\Delta \nu$ prior. Thus, the new network does not suffer from the old network's limitation.  The relatively poor performance of both old and new networks in the range $5\mu$Hz $< \Delta \nu \le 9\mu$Hz can be attributed to the difficulty in discriminating HeB secondary clump stars ($M\apprge 2.0M_{\odot}$) from RGB stars. This is because RGB and HeB stars within this $\Delta\nu$ range may have mixed mode patterns that resemble each other (\citealt[][their Figure 7]{Grosjean_2014}). 
     
\begin{figure}
		\centering
		\includegraphics[width=\linewidth]{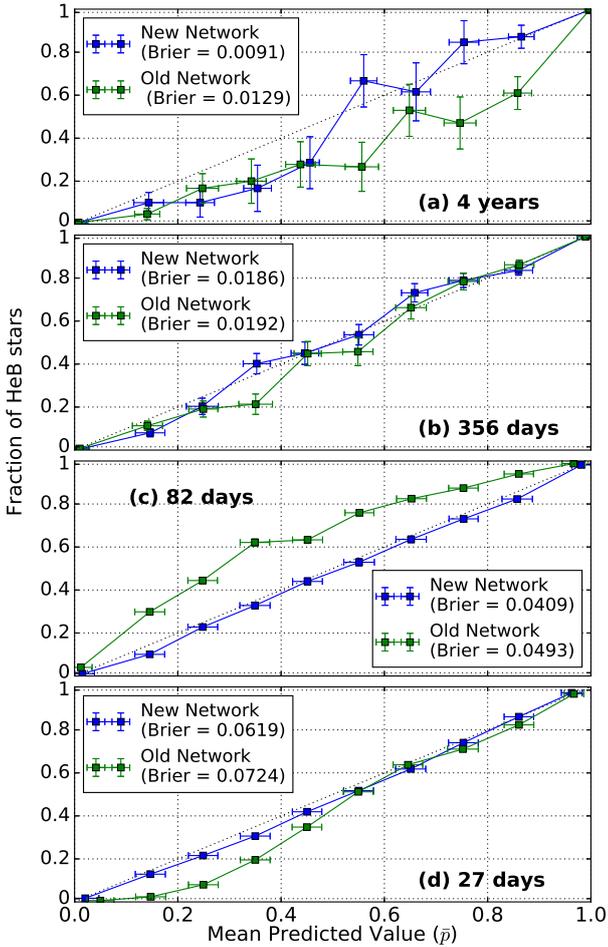}
		\caption{Reliability diagrams for (a) the 4-year classifier, (b) the 356-day classifier, (c) the 82-day classifier and (d) the 27-day classifier. The new network's probabilities are in blue, while the old network's are in green. The Brier score for each network's probabilities are listed. The dotted diagonal line represents a perfect calibration of probabilities.}
		\label{CalibrationQuad}
	\end{figure}

     Additionally, we now measure how well each network's predictions are calibrated. Each neural network classifier predicts the probability for the occurrence of each class $j$, (RGB or HeB), from the softmax function in the output layer of the network, given by:
	 
	 \begin{equation}
	 p(y=j|\mathrm{\textbf{x}}) =\frac{e^{\mathrm{\textbf{x}}\cdot\mathrm{\textbf{w}}_j}}{\sum_{k=1}^{2}e^{\mathrm{\textbf{x}}\cdot\mathrm{\textbf{w}}_k}},
	 \label{softmax}
	 \end{equation}
where $\mathrm{\textbf{x}}$ are input values to the output layer and $\mathrm{\textbf{w}}$ are the weights of the output layer (see H17 for details). Output values close to 0 indicate a high RGB probability, while values close to 1 indicate a high HeB probability. These probabilities are normalized such that they sum to 1 across both classes, and can be interpreted as the confidence level of the classifier in predicting a particular class for a star. Ideally, we would like the output probabilities to be consistent with the frequency by which we observe particular classes within a sample of stars. A classifier with such a quality is known to be \textit{well-calibrated}. For instance, if a well-calibrated classifier predicts $p = 0.6$ for each star within a sample dataset, we would expect that the true quantity of HeB stars within that sample will be around 60\%. Neural networks in general have well-calibrated probabilities \citep{Niculescu_2005}, hence we aim to confirm this for our own classifiers by observing the distribution of predictions from cross-validation and measuring the corresponding Brier score \citep{Brier_1950}, defined by:
	 
	 \begin{equation}
	 E_{\mathrm{Brier}} = \frac{1}{m} \bigg(\sum_{i=1}^{m}(y_i - \hat{y}_i)^2 \bigg),
	 \label{Brier}
	 \end{equation}
	 where $m$ is the number of stars in the dataset, $y$ the ground truth label, and $\hat{y}$ the predicted probability. Well-calibrated probabilities have $E_{\mathrm{Brier}}$ close to zero. In addition to the Brier score, we also plot a \textit{reliability diagram} \citep{Degroot_1983} by binning the classifier predictions and plotting the mean probability, $\bar{p}$, versus the true fraction of stars which are HeB within each bin. The better the calibration of the classifier, the closer the probabilities are to an increasing diagonal line from $(0,0)$ to $(1,1)$ on the plot. We show the reliability diagrams of the old and new networks in Figure \ref{CalibrationQuad}. 

	\begin{figure}
		\centering
		\includegraphics[width=\linewidth]{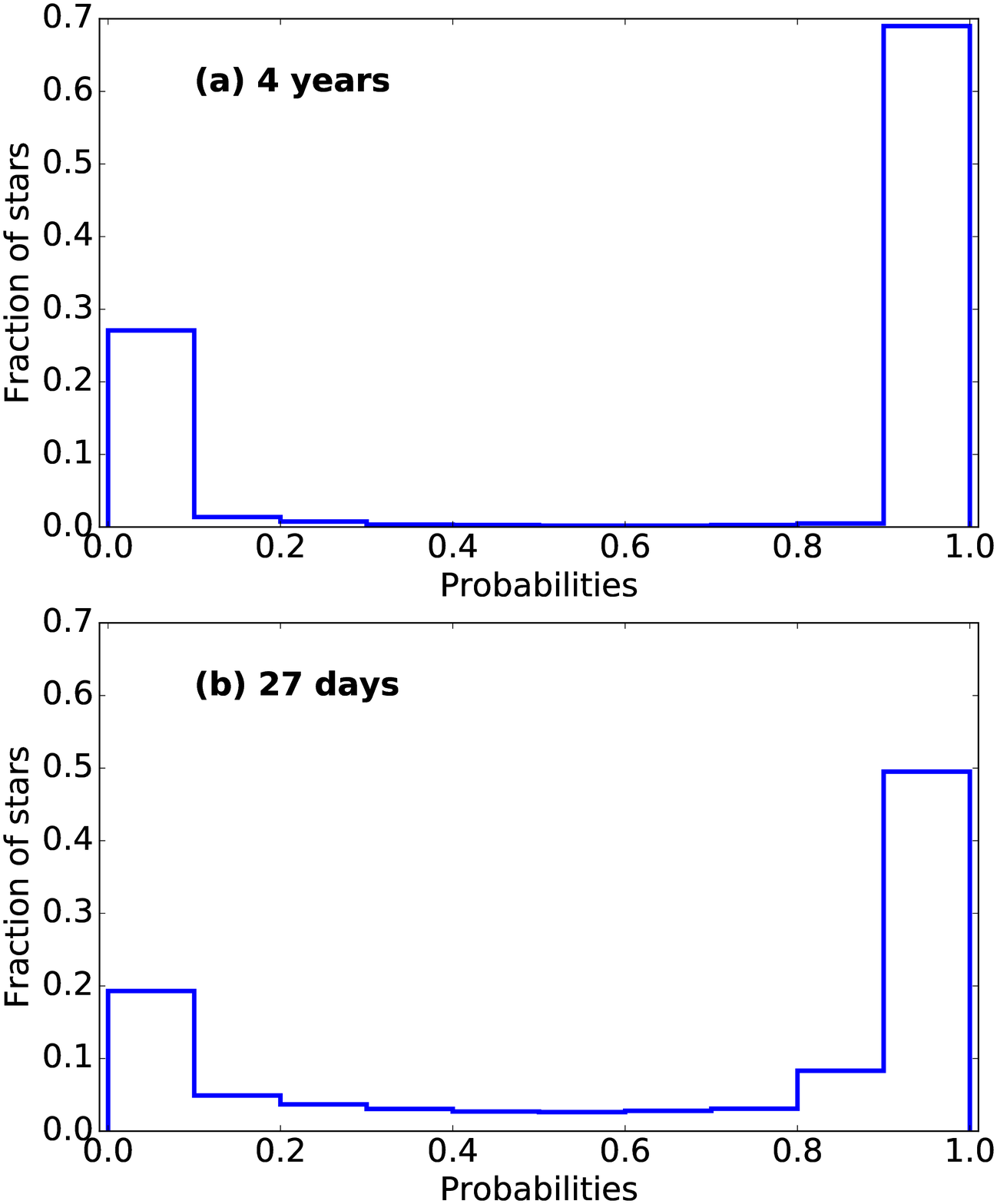}
		\caption{A typical distribution of predicted probabilities for the (a) 4-year classifier and the (b) 27-day classifier from 10-fold cross validation.}
		\label{Hist}
	\end{figure}
	We see that the new network generally has better calibrated probabilities (blue) across all data lengths compared to the old network (green), because the curves follow the diagonal line more closely and the probabilities have lower Brier scores (such as in Figure \ref{CalibrationQuad}b-d). We note that the large errorbars for intermediate probabilities ($0.2\apprle p\apprle0.8$) in Figure \ref{CalibrationQuad}a are due to the sparsity of stars with such predicted probabilities. This can be seen in Figure \ref{Hist}a, where the 4-year classifier is highly confident in its predictions, such that most stars are placed in probability bins near 0 or 1. In contrast, the classifiers for shorter data lengths do not experience the same level of sparsity for intermediate probabilities (e.g. Figure \ref{Hist}b), such that not only are the errorbars on their corresponding reliability diagrams smaller, but their curves appears to follow the diagonal line better as well (such as in Figure \ref{CalibrationQuad}d). 
	
	Through the reliability diagrams in Figure \ref{CalibrationQuad}, we can observe how the new network's predictions are biased relative to the true fraction of HeB stars. For instance, the output probabilities by the 4-year classifier using the new network will overestimate the true fraction of HeB for $0.2 \apprle \bar{p} \apprle0.5$ because the probability curve is consistently below the diagonal line for those $\bar{p}$ values. However, the probabilities for classifiers of other data lengths using the new network follow the diagonal lines well, such that no significant bias is shown across $\bar{p}$ and hence they are well calibrated.
 
From the reduced log loss values as seen in Figure \ref{LogLossSingleTrio} and better calibrated probabilities in Figure \ref{CalibrationQuad}, it is clear that the new network is an improvement over the old network. Thus, we use the new network for the classifier of each data length throughout the rest of the study. We construct our classifiers using the Keras library \citep{Keras} built on top of Theano \citep{Theano}. Training utilizes a Quadro K620 GPU and the NVIDIA cuDNN library \citep{Chetlur_2014}.  
	 
	 \subsection{Choosing a Probability Threshold}\label{Threshold}
		\begin{figure}
			\centering
			\includegraphics[width=1.01\linewidth]{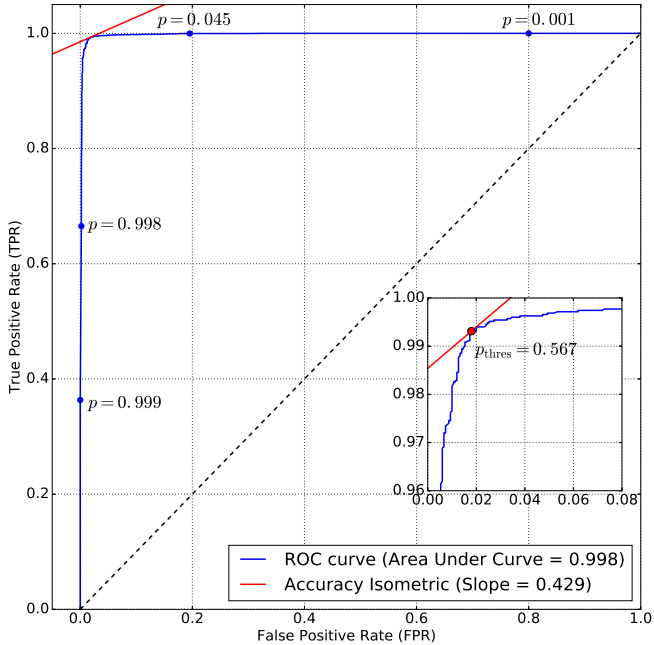}
			\caption{ROC curve of the classifier on 10-fold cross validation of 4-year \textit{Kepler} data. The black, dashed diagonal line corresponds to random guessing for predictions, while the red line is the accuracy isometric. A few points of the curve are plotted (blue dots) with their corresponding $p$ to illustrate the order of plotting. (Inset) A close-up of the plot near the point of intersection (red dot) of the accuracy isometric with the ROC curve. This point corresponds to a prediction score $p_{\mathrm{thres}}=0.567$, which is the ideal score threshold.}
			\label{ROCCurve}
		\end{figure}
		
	 Instead of using $p_{\mathrm{thres}} = 0.5$ as a fixed probability threshold for determining if a star is RGB ($p<p_{\mathrm{thres}}$) or HeB ($p\geq p_{\mathrm{thres}}$), we determine an ideal threshold that will maximize the accuracy of the classifier. We perform this by plotting the Receiver Operating Characteristic (ROC) curve \citep{Swets_2000}. As shown in Figure \ref{ROCCurve}, the ROC curve plots the true positive rate (TPR) against the false positive rate (FPR) of the classifier, defined as follows:\newline
	 \newline
	 \textbf{TPR} is the ratio of correct HeB predictions to \textit{all stars} that are truly HeB. Here it is the classifier's ability to find all HeB stars. This is also known as \textit{recall}.\newline
 	\textbf{FPR} is the ratio of stars incorrectly predicted as HeB to \textit{all stars} that are truly RGB.\newline
	 
	 The ROC curve begins at the origin in TPR-FPR space. Prediction probabilities for the evolutionary state of the stars are ranked in descending order, and this determines the order in which they are plotted on the ROC curve \citep{{Flach_2008}}. Beginning at scores close to 1, if this prediction correctly corresponds to a HeB star, we plot vertically upwards with magnitude 1/TPR. Otherwise, if the true label is actually RGB, we plot horizontally to the right with magnitude 1/FPR. Thus, a perfect classifier's ROC curve will plot all the true HeBs first (a vertical line upwards all the way to 1 on the ordinate), followed by all the true RGBs (horizontal line to the right). The ROC curve illustrates how well the classifier separates HeB stars from RGB stars, with the classifier performing better the closer the ROC curve is to the upper left corner \citep{Zweig_1993}, corresponding to an Area Under Curve (AUC) closer to a value of 1 as described in Section \ref{TestingPerformance}. Following the method described by \citet{Flach_2012}, we can use the ROC curve to determine the threshold that best separates the classes for each classifier. We first determine the \textit{accuracy isometrics} of the classifier, which are lines in TPR-FPR space with gradients equal to the ratio of RGB stars to HeB stars. We find the highest accuracy isometric that can intersect the ROC curve, which results in a single point of intersection between the accuracy isometric and the ROC curve (see Figure \ref{ROCCurve}). Because each point on the ROC curve is plotted in descending order of probability, the point of intersection between the ROC curve and the accuracy isometric corresponds to a probability, $p_{\mathrm{thres}}$, which is the best classifier probability threshold for separating RGB and HeB populations from one another.
	 
 	In Table \ref{threshtable}, we summarize the determination of the ideal probability thresholds (see Figure \ref{ROCCurve}) for separating RGB and HeB stars for all of our trained classifiers. We can interpret $p_{\mathrm{thres}}$ as follows: For a $p_{\mathrm{thres}}=0.567$, we only assign the HeB label to a star if the classifier is at least 56.7\% certain that it recognizes the input spectra to be that of a HeB star. Because we assign equal importance to predictions of both RGB and HeB stars, the ideal choice of a probability threshold should be $p_{\mathrm{thres}}=0.5$. Nonetheless, we see that we generally obtain $p_{\mathrm{thres}}$ values close to 0.5.

 	\begin{table} 
 		\centering
 		\caption{Ideal probability thresholds for classifiers corresponding to each length of data.
 		}
 		\label{threshtable}
 		\begin{threeparttable}
 			\begin{tabular}{|l|c|c|c|c|}
 				\hline
 				&\multicolumn{4}{c|}{Data Length} \\
 				\cline{2-5}
 				& 4 years & 356 days & 82 days & 27 days\\
 				\hline
 				Ideal Threshold ($p_{\mathrm{thres}}$)&0.567 &0.442 & 0.503 & 0.533\\
 				
 				\hline
 				
 			\end{tabular}
 		\end{threeparttable}
 	\end{table}
 	
 	\section{Results of the New Classifier}
	Using the methods described in the previous section, we first predict on test sets for each data length. Then we make new predictions on the `full mission' \textit{Kepler} red giants, including some stars without any prior seismic classification. Next, we apply the classifier on the K2 data of the open cluster M67. As opposed to H17, we now also determine uncertainties on the classification predictions, which we describe in Section \ref{Uncertainty}.
		\begin{table} 
			\centering
			\caption{Performance metrics for each classifier on their corresponding test set. Uncertainties are shown in compact bracket form: e.g., 0.983 (3) = $0.983 \pm 0.003$, 0.110 (17) = $0.110 \pm 0.017$.
			}
			\label{FullMetrics}
			\begin{threeparttable}
				\begin{tabular}{|l|c|c|c|c|}
					\hline
					&\multicolumn{4}{c|}{Data Length} \\
					\cline{2-5}
					& 4 years & 356 days & 82 days & 27 days\\
					\hline
					Accuracy&0.983 (3) &0.983 (3) & 0.954 (4) & 0.932 (2)\\
					Precision&0.983 (3) &0.983 (4) & 0.954 (4)&0.931 (2)\\
					Recall&0.984 (3) &0.978 (4) & 0.953 (4)&0.932 (2)\\
					F1 Score &0.984 (2) &0.978 (4)&0.953 (4)&0.932 (2)\\
					ROC AUC &0.998 (1) &0.996 (1)&0.986 (1)&0.981 (1)\\
					Log Loss &0.052 (6) &0.056 (5)&0.137 (9)&0.170 (3)\\
					Brier & 0.013 (2) & 0.014 (2)& 0.036 (2)&0.049 (1)\\
					\hline
					
				\end{tabular}
			\end{threeparttable}
		\end{table}
		\begin{table*} 
			\centering
			\caption{Evolutionary state classifications for 14983 \textit{Kepler} red giants (0 - RGB, 1 - HeB), with probability $p$ of the star being HeB and its corresponding uncertainty $\sigma_p$. These classifications use $p_{\mathrm{thres}}=0.567$, as discussed in Section \ref{Threshold}. Listed values of $\Delta\nu$ and $\nu_{\mathrm{max}}$ are from \citet{Yu_2018}, while $\epsilon$ values are obtained from the method described in \citet{Stello_2016a}. The flag `U' indicates that the star is in the previously unclassified set, while flags 'R' and 'T' indicate whether the star is present in the training or test sets, respectively. Additional flags indicate whether the predicted label is disputed by classifications by \citet{Stello2013} (`S') , \citet{Elsworth2016} (`E'), or \citet{Mosser2014}/\citet{Vrard2016} (`V'). Values of $-99$ for $\epsilon$ indicate values that are not provided. The full version of this table is available in a machine-readable form in the online journal, with a portion shown here for guidance regarding its form and content.
			}
			\label{SampleTable}
			\begin{threeparttable}
				\begin{tabular}{|c|c|r|r|r|r|r|c|}
					\hline
					KIC & Classification & $p$ \hspace{0.25cm} & $\sigma_p$\hspace{0.25cm} & $\Delta\nu$ ($\mu$Hz) & $\nu_{\mathrm{max}}$ ($\mu$Hz) & $\epsilon$ \hspace{0.25cm} & Flag\\
					\hline
					4058403&1&0.978&0.024&5.25476&51.89827&0.705&TV\\
					4284099&1&0.990&0.011&5.15982&56.61160&1.005&T\\
					4284170&0 &0.067 & 0.120 & 6.18558 & 62.28761 & 1.080 & U \\
					4284281&0 &0.120& 0.123&4.41799 & 45.47096 &1.055 & U \\
					4284367&1 &0.999 & 0.001&4.31112 & 36.86113 & 0.875 & R\\
					4284581 &0 &0.428&0.373&4.70426&44.89361 & 0.555 & UES\\
					4284739 &1 &0.999&0.001&4.27640&34.09092&0.895& U\\
					4345370 & 0 & 0.144& 0.155&4.04844&32.68148&1.030&U\\
					4345809&1&0.976 & 0.087 &3.72018&31.28668&1.030&U\\
					4346339&0&0.000&0.001&14.49812&195.42500&-99.000&U\\
					...&...&...&...&...&...&...&...\\
					\hline
					
				\end{tabular}
			\end{threeparttable}
		\end{table*}
	\subsection{Uncertainties for Classifier Predictions}\label{Uncertainty}
	When predicting on test sets or on real datasets, we cannot use the standard deviation across cross-validation folds as our uncertainty. Instead, we measure the uncertainty of each prediction produced by the classifiers using Monte Carlo dropout to approximate a Gaussian process \citep{Gal_2016}. Dropout is an effective way to prevent overfitting by randomly setting neurons within the neural network to zero with a certain probability \citep{Srivastava_14}. It is often implemented differently when the classifier is being trained as opposed to when it is predicting on test data. During training, dropout is implemented by randomly setting neurons to zero with a certain drop out probability, but during testing, it is implemented by scaling the weights for these neurons with the value of the dropout probability.  By passing data through the network multiple times during testing with the `training' form of dropout enabled, we effectively perform a Monte Carlo integration over a Gaussian process posterior approximation \citep{Gal_2016}. Hence, we report classifier predictions as the expectation of 50 passes through the network, with the standard deviation as the prediction uncertainty.

	\subsection{Test Set Performance}\label{TestPerformance}
	
	The performance of the classifiers on their respective test sets are shown in Table \ref{FullMetrics}. In general, the shorter the data length, the lower the classifier performance. This is expected because the folded spectra of shorter data lengths have lower frequency resolution. A notable exception to this is the performance for the 356-day classifier. This shows that 356-day data still have the level of detail as the 4-year data for the purpose of classification (see Figure \ref{DifferentTimes}). 
	\subsection{Revised \textit{Kepler} Unclassified Set Predictions}	
		\begin{figure}
			\centering
			\includegraphics[width=1.05\linewidth]{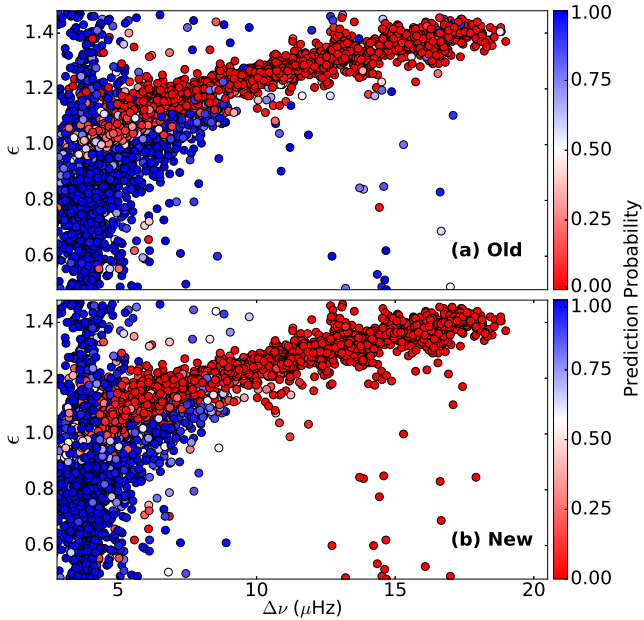}
			\caption{Predictions on the unclassified set of \textit{Kepler} red giants. (a) Previously given classifications for 7655 red giants by \citet{Hon_2017}, with the old network. (b) Our new classifications for 8633 red giants, produced by a convolutional neural network classifier with the new network structure. 7655 from these are revised classifications for (a). The colour bar reflects the change in the probability threshold ($p_{\mathrm{thres}}=0.567$).}
			\label{OldNew}
		\end{figure}
	We provide new or updated predictions for 8633 \textit{Kepler} red giants with $\Delta\nu>2.8\mu$Hz and no prior asymptotic period spacing measurements, from which 7655 were already classified in H17 as part of the \textit{unclassified set}. This new set of 8633 red giants contains 426 red giants that have not been given any seismic-based evolutionary state classifications prior to this study (not even by H17). We present our results in a comprehensive table (available online), which includes evolutionary state classifications for the 6350 stars in our training and test sets, hence a total of 14983 stars are classified. The classifications for the training set are obtained by 10-fold cross validation. In particular, they are predictions for the stars in the validation fold for each iteration of the 10-fold cross validation. A sample of this catalogue is shown in Table \ref{SampleTable}.
	
	From a comparison of old (H17) and new (this work) classifications on the unclassified set in Figure \ref{OldNew}, it can be seen that our new network now correctly does not predict any HeB stars at $\Delta\nu > 10\mu$Hz. Furthermore, stars with $\epsilon$ and $\Delta\nu$ that appear to follow the $\Delta\nu - \epsilon$ relation of RGB stars \citep{Kallinger_2012} are also now more likely to be predicted as RGB stars. Interestingly, there is still a small population of predicted RGB stars at $\Delta\nu \simeq 5-6\mu$Hz with $\epsilon<0.8$. It may be worthwhile in future work to investigate these stars that appear to deviate from the RGB $\Delta\nu - \epsilon$ relation.  
		
	\subsection{M67 Red Giant Predictions}
	\begin{table} 
		\centering
		\caption{Predictions on M67 giants from K2 data with $\Delta\nu>2.8\mu$Hz using the 82-day classifier ($p_{\mathrm{thres}} = 0.503$), with classifier probabilities $p$. Values for $\Delta\nu$, $\nu_{\mathrm{max}}$, and colour-magnitude diagram (CMD) classifications are obtained from \citet{Stello_2016}. Disputes are highlighted in bold.
		}
		\label{M67Table}
		\begin{threeparttable}
			\begin{tabular}{|c|c|c|c|r|c|}
				\hline
				EPIC ID & $p (1\sigma)$ & Classifier& CMD & $\nu_{\mathrm{max}}$ & $\Delta\nu$ \\
				\hline
				211406540&0.994 (0.004) &HeB&HeB & 31.8&4.59\\
				\textbf{211406541} & \textbf{0.766 (0.205)} & \textbf{HeB}&\textbf{RGB}&\textbf{34.7}&\textbf{4.27}\\
				211418433 & 0.995 (0.005) & HeB & HeB & 36.5 & 4.18\\
				211410523&0.843 (0.172) &HeB & HeB &37.8 & 3.95\\
				211415732&0.994 (0.010) & HeB & HeB & 37.9 & 4.40 \\
				211420284& 0.959 (0.032) & HeB & HeB & 39.3 & 4.22 \\
				211413402&0.593 (0.320) & HeB & HeB & 39.5 & 4.44\\
				211417056 & 0.943 (0.077) & HeB & HeB & 39.6 & 4.17 \\
				211392837 & 0.103 (0.107) & RGB & RGB & 47.5 & 4.81\\
				211413623 & 0.012 (0.016) & RGB & RGB & 64.8 & 6.28 \\
				211396385 & 0.006 (0.008) & RGB & RGB & 77.4 & 7.00 \\
				211414300 & 0.307 (0.250) & HeB & RGB & 78.8 & 7.19 \\
				211408346 & 0.070 (0.062) & RGB & RGB & 98.7 & 8.17 \\
				211410231 & 0.048 (0.053) & RGB & RGB & 103.1 & 8.87 \\
				211412928 & 0.001 (0.002) & RGB & RGB & 117.8 & 9.74 \\
				211411629 & 0.001 (0.001) & RGB & RGB & 196.0 & 14.43 \\
				211414687 & 0.000 (0.001) & RGB & RGB & 203.0 & 15.10\\
				211416749 & 0.000 (0.001) & RGB & RGB & 234.3 & 16.76 \\
				211421954 & 0.000 (0.001) & RGB & RGB & 246.1 & 17.47 \\
				211409560\tnote{a} & 0.000 (0.001) & RGB & RGB & 272.2 & 19.10\\
				211388537\tnote{a} & 0.000 (0.001) & RGB & RGB & 287.6 & 20.15\\
				211403248\tnote{a} & 0.000 (0.001) & RGB & RGB & 305.5 & 21.45\\	
				211415364\tnote{a} & 0.000 (0.001) & RGB & RGB & 463.0 & 28.29 \\
				211411922\tnote{a} & 0.000 (0.001) & RGB & RGB & 559.0 & 36.34 \\
				211409088\tnote{a} & 0.000 (0.001) & RGB & RGB & 562.0 & 33.02 \\
				\hline
				
			\end{tabular}
			\begin{tablenotes}
				\item[a] We use short cadence data for this star.
			\end{tablenotes}
		\end{threeparttable}
	\end{table}
	
	To test our 82-day classifier on real K2 data, we obtain the background-corrected K2 spectra along with measurements of $\Delta\nu$ and $\nu_\mathrm{max}$ of red giants within the open cluster M67 from the study by \citet{Stello_2016}. From the cluster, we predict the evolutionary state of 25 red giants with $\Delta\nu > 2.8 \mu$Hz, following the $\Delta\nu$ range limitation of our classifier. We compare our asteroseismically-derived classifications with the evolutionary state classifications from the colour-magnitude diagram (CMD). The outcome is tabulated in Table \ref{M67Table}. Our classifications are in good agreement with those inferred from the colour-magnitude diagram (CMD). Only one red giant has a disputed classification, where the classifier believes the star is HeB. The high level of agreement with CMD classifications indicate a good classifier performance.
	
	Reporting the uncertainties of each prediction shows which stars are the most difficult to classify. Large uncertainties imply the classifier is highly uncertain about its degree of belief. If the resulting $p$ is close to the 82-day probability threshold $p_{\mathrm{thres}}=0.503$, this then implies that the classifier is highly uncertain on the resulting predicted class. In Table \ref{M67Table}, we identify the disputed star EPIC 211406541 showing such an uncertain prediction. Interestingly, we see that the classifier is able to correctly predict the evolutionary states for RGB stars with $\nu_{\mathrm{max}} > 280\mu$Hz despite only having trained on long cadence data. This is because the additional $\Delta\nu$ input (Figure \ref{AdditionModel}) allows the classifier to be able to generalize its predictions to red giants above the long cadence Nyquist frequency ($\nu_{\mathrm{Nyq}}\simeq280\mu$Hz).
	
		\begin{figure}
			\centering
			\includegraphics[width=\linewidth]{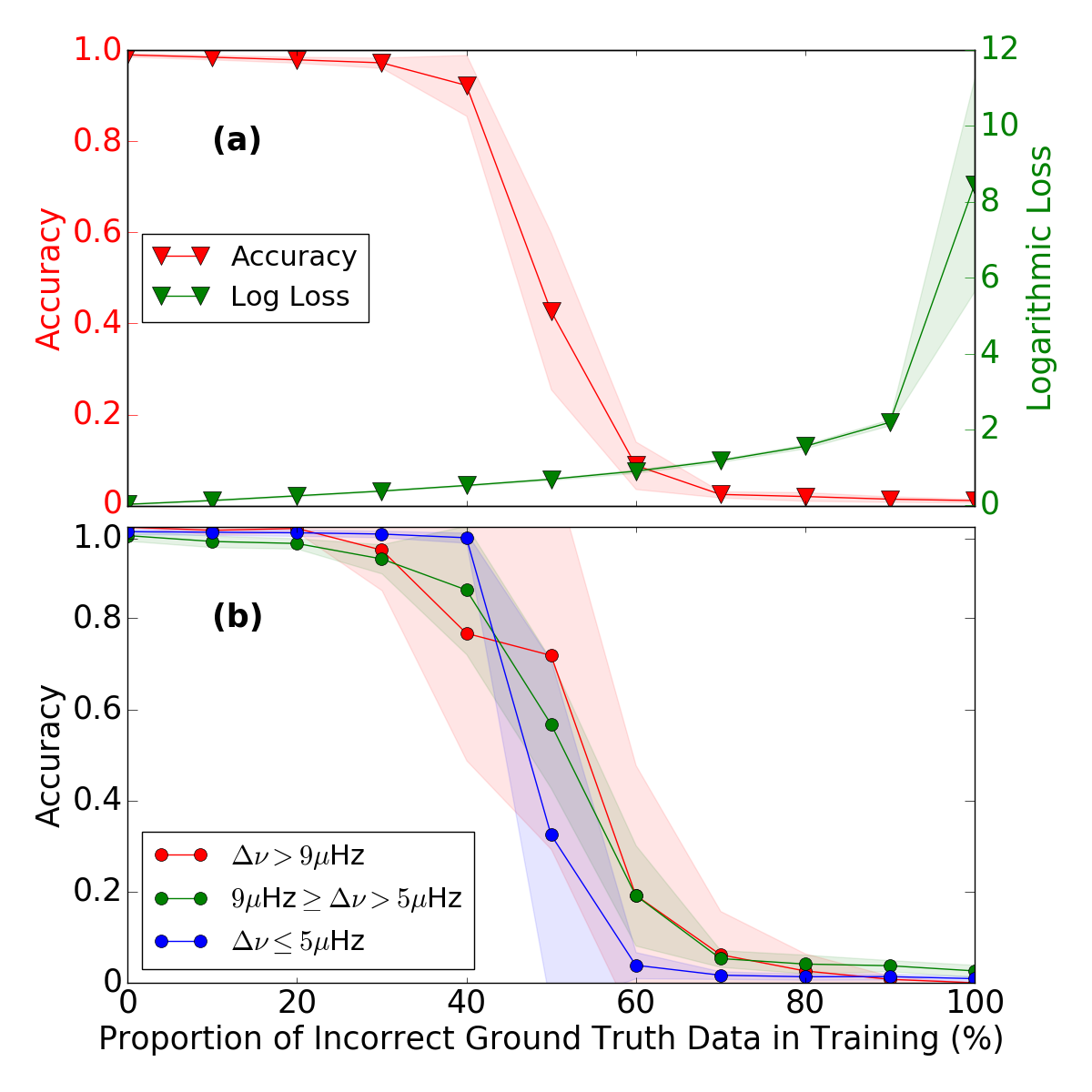}
			\caption{(a) Accuracy and log loss of the 4-year classifier when trained on data with a proportion incorrect labels. Evaluation was performed using 10-fold cross validation. (b) The decomposition of the accuracy curve in (a). The shaded regions are $1\sigma$ uncertainty ranges.}
			\label{FlipGraph}
		\end{figure}
	\section{Robustness of the New Classifier}
	To test the robustness of our deep learning classifiers, we determine the impact of suboptimal training or testing conditions on classifier performance.
	
	\subsection{Flipping the Truth Labels of Training Data}\label{FlipLabels}	
	
	\begin{figure}
		\centering
		\includegraphics[width=\linewidth]{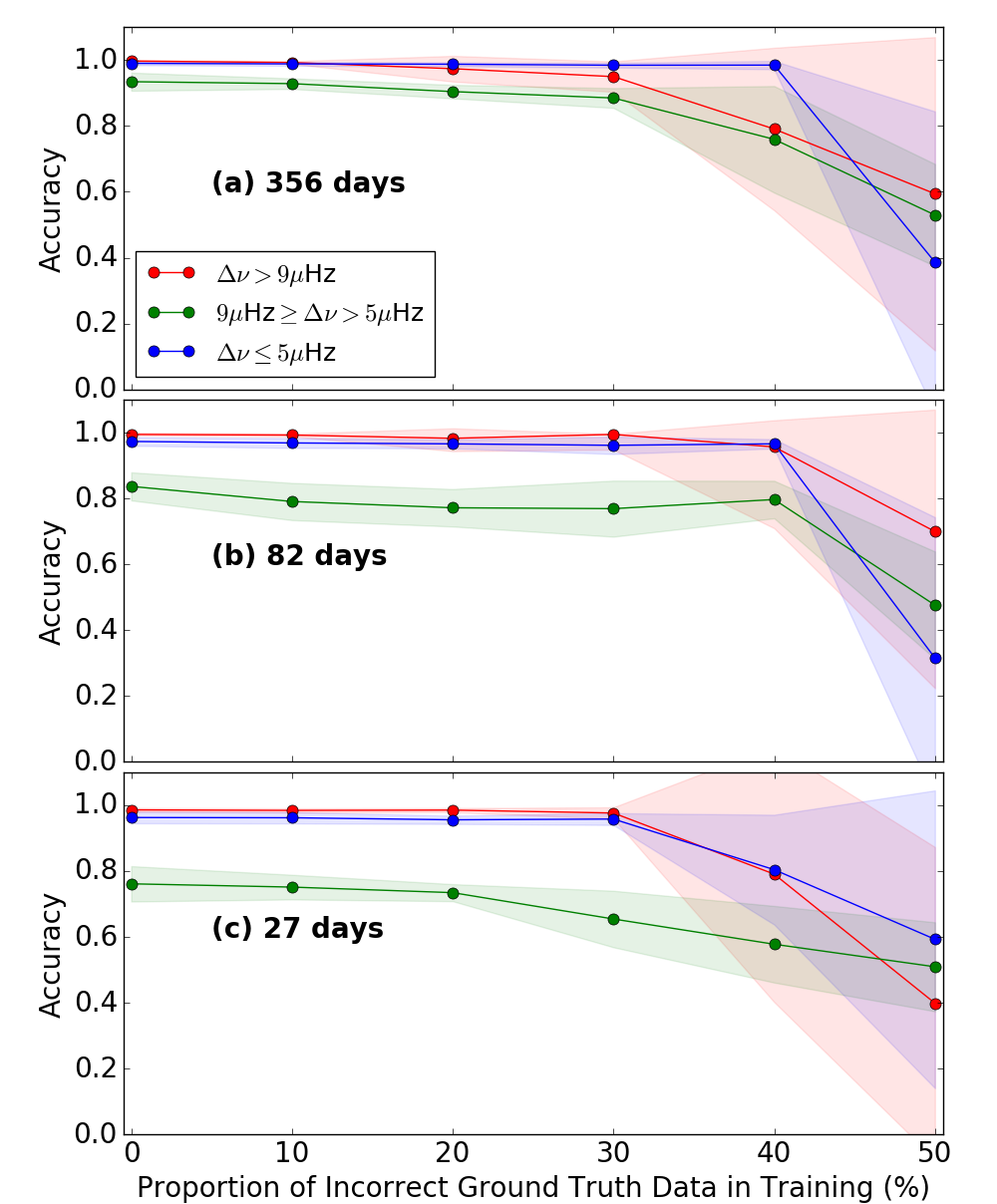}
		\caption{The decomposition of accuracy curves for (a) 356-day, (b) 82-day, and (c) 27-day data as a function of the number of incorrect training data by 10-fold cross validation. The shaded regions are $1\sigma$ uncertainty ranges. To be concise, proportions of flipped labels only up to 50\% are plotted due to the symmetry of the accuracy curves about this ordinate.}
		\label{FlipGraph2}
	\end{figure}
	\begin{figure*}
		\centering
		\includegraphics[width=0.8\linewidth]{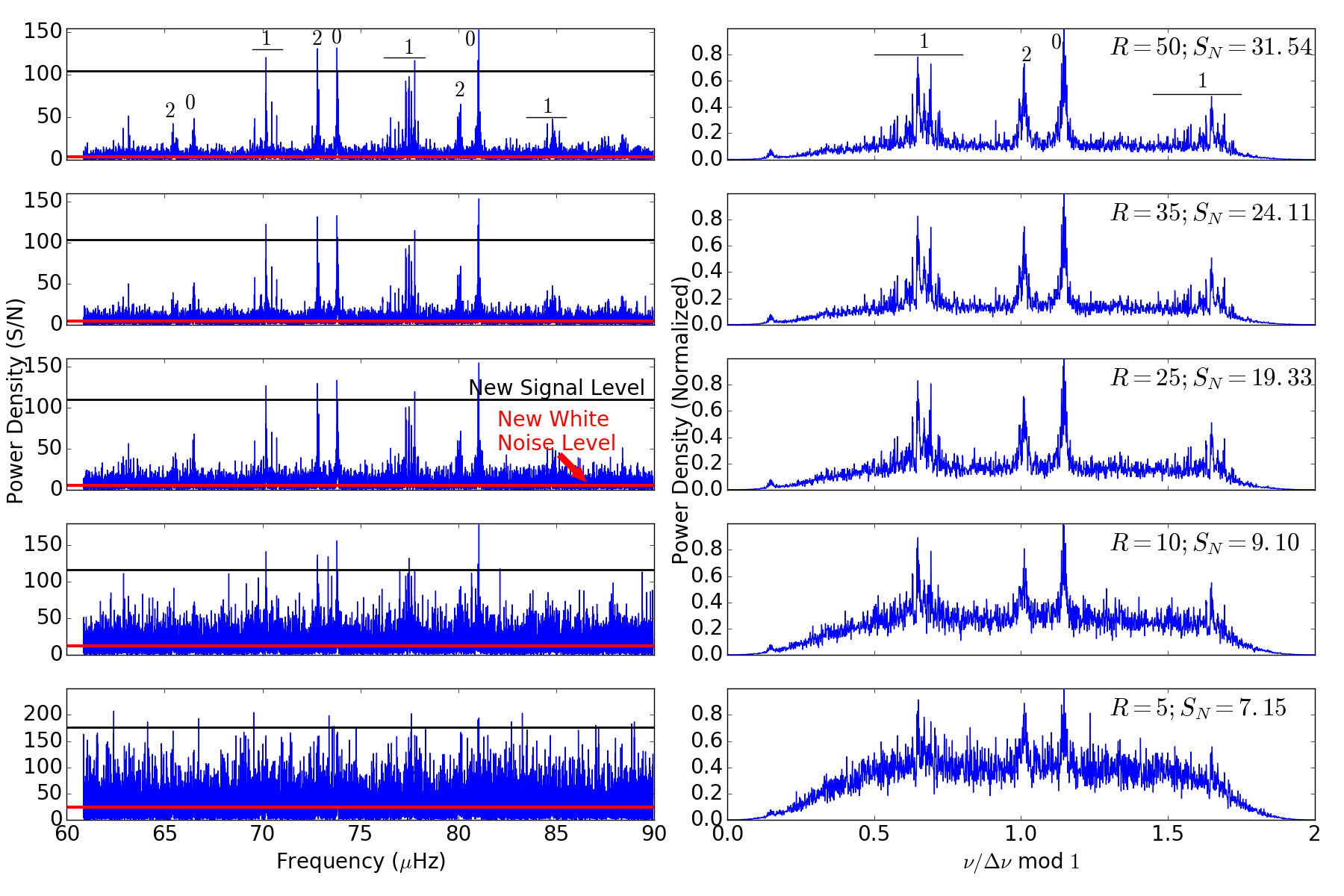}
		\caption{Noise normalised power spectra (left) and folded spectra (right) of 4-year \textit{Kepler} data of RGB star KIC 8144802, after adding white noise. The resulting quality of the spectra is quantified by a measured signal-to-noise ratio, $S_N$, which determines how many times the signal level (black line) is greater than the noise level (red line). The amount of added noise to the original spectra is defined as the signal level \textit{prior} to adding noise, divided by the factor $R$.}
		\label{NoiseFigure}
	\end{figure*}
	
	We aim to simulate the scenario in which assigned training labels may be unreliable by randomly flipping population labels of stars in the training set from HeB to RGB and vice versa during the process of training the classifier. The classifier performance is measured using 10-fold cross validation because training with suboptimal data can cause large variations in performance. We randomly flip a percentage of training labels during training, from 0\% to 100\% in 10\% increments.We then make predictions on the validation fold, which does not have any stars with flipped labels. This analysis illustrates the dependence of the classifier performance on the accuracy of its ground truth during training.
	
	Figure \ref{FlipGraph}a shows the behaviour of the classifier trained on 4-year data as we introduce incorrectly labelled, or label-flipped,  ground truth into the training set. The classifier is remarkably robust, with the accuracy remaining almost constant until about 40\% incorrect training labels. Thus, the predictions are generally resistant to wrong information learned during training as long as the majority of training examples are correct. Meanwhile, the log loss shows the expected monotonically increasing behaviour indicative of an increase in uncertainty as more examples in the training set are incorrect. The very high log loss at a 100\% of incorrect labels in the training set indicates that the classifier is now confidently giving very incorrect predictions.
	
	By examining the decomposition of the classifier's accuracy according to $\Delta\nu$ regions as defined in Figure \ref{LogLossSingleTrio}, we can reveal interesting prediction behaviours of the classifier. We show this for the 4-year classifier in Figure \ref{FlipGraph}b. We see that all curves are approximately symmetrical about the 50\% flipped label ordinate, with different $\Delta\nu$ regions having different accuracies and uncertainty ranges as the proportion of incorrect training data varies. Because the accuracy and log loss behaviours (as of Figure \ref{FlipGraph}a) for the classifiers of shorter data lengths are similar to that of the 4-year classifier, we do not show them. However, we do present their accuracy curve decompositions in Figure \ref{FlipGraph2}. Due to the symmetry of the curves, we only plot up to 50\% of flipped training labels. 

	Similar to the 4-year classifier in Figure \ref{FlipGraph}b, the 356-day, 82-day, and 27-day classifiers require at least $\simeq$ 40\% of incorrect training data before they experience a significant impact in performance (drop in accuracy greater than 10\%). Next, we see that although the curve for $\Delta\nu > 9\mu$Hz (red) for all classifiers has the highest accuracy for low proportions of incorrect training data, it develops a high level of uncertainty (broad red band) from about a 30\% proportion of incorrect training data onwards. In this $\Delta\nu$ range, the training set mainly contains RGB stars. Providing a fraction of incorrect training data in this $\Delta\nu$ range forces the classifier to learn a rule that allows it to split very similar images with similar $\Delta\nu$ into two separate classes. The difficulty of this induces a large degree of uncertainty in predictions. In contrast, for other regions where both true HeB and RGB stars exist, it may be easier for the classifier to learn a rule to split the data into two classes as long as there exists a significant proportion of training data that are assigned different classes and are not too identical to one another. In support of this, we see from Figures \ref{FlipGraph} and \ref{FlipGraph2} that the accuracy curves for $\Delta\nu \leq5\mu$Hz (blue) generally only develop high uncertainties near a 50\% proportion of flipped training labels. We also note that across all classifiers, the accuracy curves for $5\mu$Hz $< \Delta \nu \le 9\mu$Hz (green) are consistently the lowest in accuracy for low proportions of flipped training labels, which agrees with our result in Figure \ref{LogLossSingleTrio}. Although this is the case, we see that the green curves do not develop very large uncertainties as the proportion of incorrect training data increases. It is likely that this is related to the intrinsic difficulty of classifying RGB and HeB stars at that range of $\Delta\nu$.

	\subsection{Added White Noise in Test Data}\label{WhiteNoise}

	\begin{figure}
	\centering
	\includegraphics[width=0.965\linewidth]{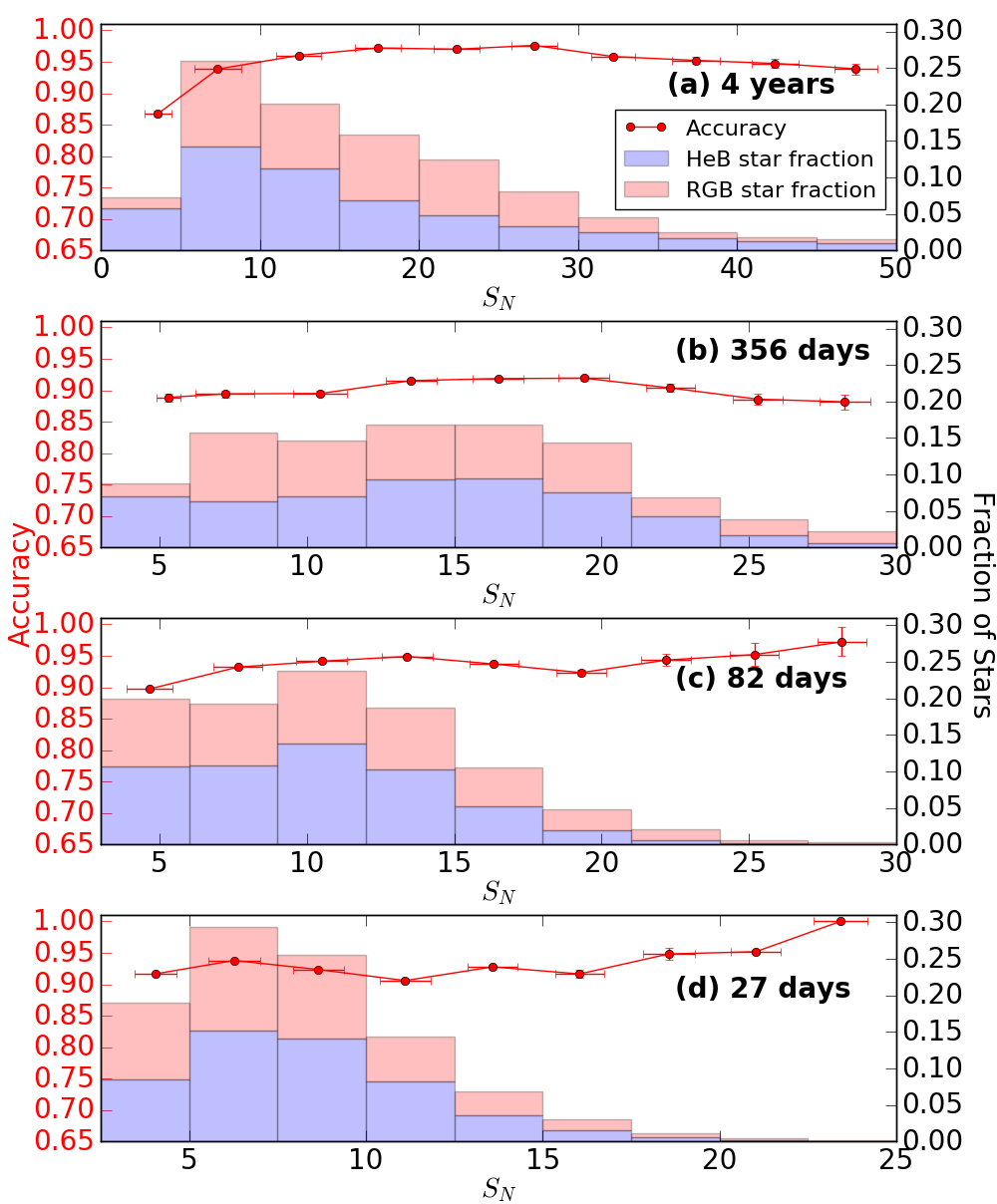}
	\caption{Accuracy of each trained classifier on test sets with added white noise, as a function of the resulting observed signal-to-noise ratio, $S_N$. The histograms in each plot show the fraction of test stars within each $S_N$ bin.}
	\label{NoiseTestGraph}
	\end{figure}
		
	 In previous sections, we have assumed that the test set is of similar quality to the training set. In reality, this is not always the case, because future datasets may be very different from \textit{Kepler} data in terms of quality, though we would still want to use \textit{Kepler} data to train classifiers for K2 and TESS data for instance. This is possible because recent K2 light curves have comparable photometric precision to \textit{Kepler} light curves \citep{Luger_2016}, while TESS light curves are expected to have similar properties as to those from \textit{Kepler}, but for stars 4-5 magnitudes brighter due to its smaller aperture \citep{TESS}. Hence, we can analyse how well the classifier perfoms at lower signal-to-noise ratios for K2- and TESS-like power spectra. This is done by studying the tolerance of the classifier performance when predicting on spectra in the test set with 'distortions' in the form of added white noise.
	
	We use power spectra divided by the background noise, such that the mean noise level is 1.0. To simulate additional white noise, we add specific levels of noise sampled from a $\chi^2_2$ distribution onto the entire power spectrum.   We first define the signal level of the power excess as the median power of the 30 highest power frequency bins within a $4\Delta\nu$ range around $\nu_{\mathrm{max}}$. We control the mean noise level that will be added to the spectrum by dividing this median signal by a factor $R$. As an example, using $R=1$, the mean noise level that will be added is the same as the initial signal level. We then add this mean noise level to the spectra and quantify the resulting spectra quality using a signal-to-noise ratio, $S_N$, which is the \textit{new} signal level of the spectrum, divided by the \textit{new} mean noise level. Thus, by definition, $S_N$ is similar to $R$, except that it is measured \textit{after} adding white noise to the spectra (see Figure \ref{NoiseFigure}). We use $R$ with values of 1, 5, 10, 15, up to 50 in steps of 5 to generate noisier versions of the test sets. We then test the classifier performance on these noisier test sets and plot the performance versus binned $S_N$ values in Figure \ref{NoiseTestGraph}. 

	We see that the classifiers are very robust towards noise by  maintaining accuracies above 90\% across all ranges of tested $S_N$ levels. The only exception to this is for $S_N \apprle 5$ for the 4-year classifier in Figure \ref{NoiseTestGraph}a. As seen in Figure \ref{NoiseFigure}, this range of $S_N$ is where the white noise level starts obscuring oscillation modes of lower amplitude in both the power spectrum and the folded spectrum. Classification at such low $S_N$ is thus difficult for both the expert eye and the machine expert. Nonetheless, we partially attribute this overall robustness towards white noise levels to the use of folded spectra. Because oscillation modes in the folded spectra generally overlap one another, they can still be distinctly observed even in the presence of large amounts of white noise (see lower rows in Figure \ref{NoiseFigure}). Hence, the classifiers can still detect them even on data with lower frequency resolution and can still make reasonably accurate predictions at low $S_N$ values as seen in Figures \ref{NoiseTestGraph}c and \ref{NoiseTestGraph}d. Interestingly, we see that the 4-year and 356-day classifiers do not achieve the test set accuracies in Table \ref{FullMetrics} in the limit of high $S_N$. One possible explanation is that a significant number of stars that are difficult to classify have been placed into such $S_N$ bins. An alternate possibility is the presence of \textit{adversarial examples} \citep{Szegedy_13}, where small perturbations in the form of artificial noise added to a few pixels on a standard image produces an image highly identical to the original visually, but is able to trick a deep learning image classifier to recognize this perturbed image very differently \citep{Nguyen_2014}. The investigation of this is currently beyond the scope of this paper.
	
	\subsection{$\Delta\nu$ Variations}\label{PerturbDnuSection}
	While we have accounted for less precise $\Delta\nu$ values in Section \ref{DnuPrecision}, it is still insightful to investigate how the classifiers perform on stars with $\Delta\nu$ values that deviate by a fixed fraction of their `correct' 4-year value. Doing this provides us with a measure of classifier robustness against varying $\Delta\nu$ uncertainties. Furthermore, this allows us to investigate the forward compatibility of the classifier with respect to $\Delta\nu$, such that we can form a reasonable estimate of how much better or worse the classifiers can perform when predicting on a dataset with better or worse $\Delta\nu$ uncertainties than that shown in Figure \ref{DNUFractionalError}.
		
\begin{figure}
	\centering
	\includegraphics[width=\linewidth]{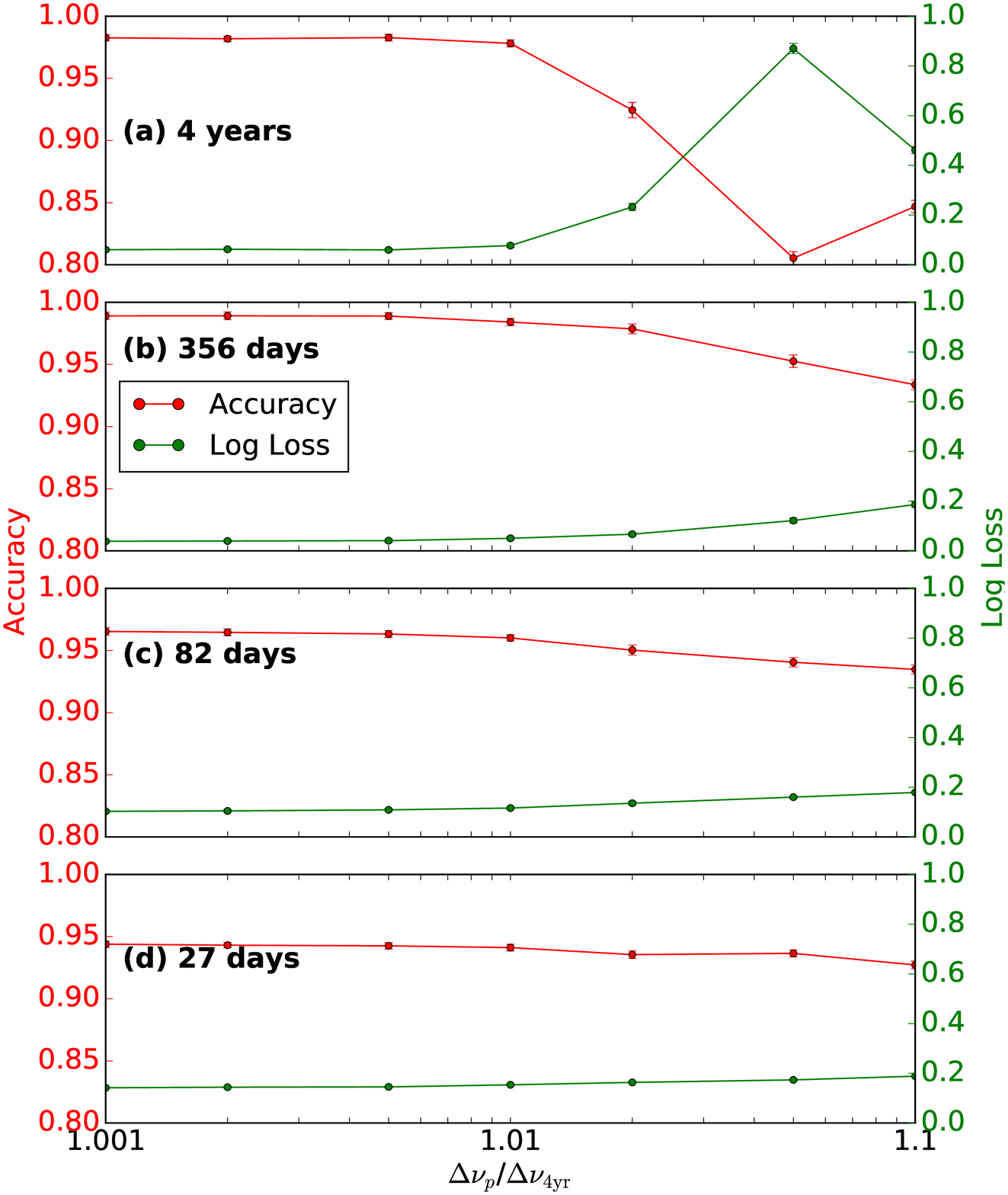}
	\caption{Classifier performance on test sets constructed using $\Delta\nu_p$, which are $\Delta\nu$ that are perturbed from $\Delta\nu_{\mathrm{4yr}}$ by a fixed fraction of that value.}
	\label{TestDnu}
	\end{figure}

	We first perturb the 4-year $\Delta\nu$ of each star, $\Delta\nu_{\mathrm{4yr}}$, in the test set by a fixed fraction of its value, and denote the resulting value as $\Delta\nu_p$. While the perturbation magnitude is fixed, its direction is random, such that the perturbation is either added or subtracted. We perturb copies of test sets with $\Delta\nu$ fractional differences of 0.1\%, 0.2\%, 0.5\%, 1\%, 2\%, 5\%, and 10\%. We note that while the 356-day, 82-day, and 27-day test sets that we have previously constructed in this study have used less precise $\Delta\nu$ values with uncertainty distributions shown in Figure \ref{DNUFractionalError}, the corresponding test sets that we construct in this Section initially use $\Delta\nu_{\mathrm{4yr}}$ that we then perturb. We plot the classifier performance on these perturbed test sets in Figure \ref{TestDnu}. 
	\begin{figure}
		\centering
		\includegraphics[width=1.02\linewidth]{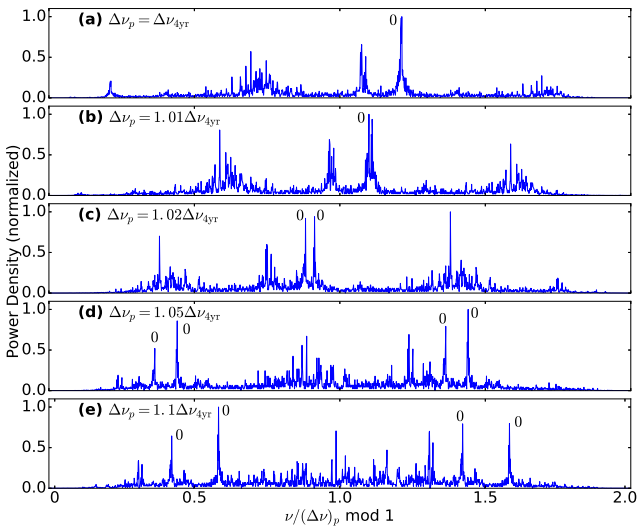}
		\caption{4-year folded spectra of KIC 3641504 ($\Delta\nu \simeq 8.857 \mu$Hz), made using a perturbed frequency spacing, $\Delta\nu_p$. The position of the main $l=0$ mode in panel (a) is tracked from panels (a) to (e) for clarity. As $\Delta\nu_p$ becomes more perturbed from $\Delta\nu_{\mathrm{4yr}}$, oscillation modes in the image are systematically translated horizontally. From panel (c) onwards, the $l=0$ modes no longer overlap and scatter across the image, with recurring instances on the right in panels (d) and (e) by wrapping around.}
		\label{DnuFrac}
	\end{figure}
	As expected, more perturbed $\Delta\nu$ values result in worse classifier performances. We show the effect of perturbed $\Delta\nu$ values on 4-year folded spectra in Figure \ref{DnuFrac}. From the Figure, we see that for $\Delta\nu_p \apprge 1.01\Delta\nu_{\mathrm{4yr}}$, oscillation modes (features) of the same spherical degree, $l$, do not overlap each other well, causing a more complicated mode distribution in the folded spectrum, which potentially confuses the classifier. This agrees with the significant drop in performance for the 4-year classifier in Figure \ref{TestDnu}a at $\Delta\nu_p \apprge 1.01\Delta\nu_{\mathrm{4yr}}$. However, we see in Figures \ref{TestDnu}b-d that classifiers of shorter data lengths are much more robust towards $\Delta\nu$ perturbations than the 4-year classifier, because their accuracies do not decrease as easily with $\Delta\nu_p$. In particular, the shorter the data length, the smaller the effect of higher $\Delta\nu$ perturbations towards the classifier performance. There are two reasons for this, with the first being the lower frequency resolution of shorter data lengths, which makes the effect of small $\Delta\nu$ perturbations less pronounced visually on their folded spectra than on the 4-year-based folded spectra. Secondly, the classifiers of shorter data lengths are more likely to have `seen' folded spectra made with less precise $\Delta\nu$ values because they trained on data using $\Delta\nu$ with higher uncertainties than the 4-year $\Delta\nu$ measurements. In contrast, the 4-year classifier has trained using $\Delta\nu$ values with small uncertainties and thus experiences difficulties generalizing to $\Delta\nu$ values with a large uncertainty.
	
	When the adopted $\Delta\nu$ is close to $\Delta\nu_{\mathrm{4yr}}$, the 356-day, 82-day, and 27-day classifiers notably achieve higher test accuracies (98.9\%, 96.5\%, and 94.3\%, respectively) than their test accuracies listed in Table \ref{FullMetrics}, which were based on more uncertain, but more realistic $\Delta\nu$ values. This suggests that these classifiers are using the right features for classification because they can better recognize features that overlap well in the folded spectra. Interestingly, we note that the accuracy for the 4-year classifier does not monotonically decrease (or increase for log loss) with increasing $\Delta\nu_p$. At $\Delta\nu_p\simeq1.1\Delta\nu_{\mathrm{4yr}}$, the classifier accuracy increases (or decreases for log loss), which may be due to the overlapping of certain modes, such as $l=0$ with $l=2$, at particular fractional $\Delta\nu$ differences being correctly recognized by the classifiers.

	\section{Conclusions}
	We have developed a variant of our previous deep learning classifier, which improves the prediction accuracy of our previous classifier and avoids predicting HeB stars at high $\Delta\nu$. We trained improved classifiers for 4-year, 356-day, 82-day, and 27-day photometric time series, which are representative for \textit{Kepler}, K2, TESS, and large parts of the PLATO sample. In order to optimize the classifier's performance, we determined the probability threshold $p_{\mathrm{thres}}$ that maximizes the classifier accuracy. As a result, we could report test set accuracies of 98.3\% (4 years), 98.3\% (356 days), 95.4\% (82 days), and 93.2\% (27 days).
	
	Next, we presented the evolutionary state classifications of 14983 \textit{Kepler} red giants using the 4-year classifier, which comprised 5000 training stars, 1350 test stars, and 8633 stars without asymptotic period spacing measurements. The classifications for these 8633 stars include 426 that previously had no seismic classification at all. The availability of evolutionary state classifications for this many \textit{Kepler} red giants will be a critical foundation in further expanding joint asteroseismic and spectroscopic investigations focused on improving our understanding in stellar astrophysics and galactic archaeology (e.g. \citealt{Pinsonneault_2014, Constantino_2015, Aguirre_2018}). Besides \textit{Kepler} red giants, we also predicted the evolutionary states of red giants within the open cluster M67 using the 82-day classifier on K2 data. The predictions showed good agreement with classifications derived using the colour-magnitude diagram, with only one disputed prediction.
	
	We then analysed the robustness of the classifiers, and found that they are remarkably robust towards incorrect ground truths, needing approximately 40\% of training data to be incorrect before the classifier performance is significantly affected. We also found that the classifiers can still make accurate predictions in the presence of significant white noise levels. By testing the classifiers on folded spectra constructed using perturbed $\Delta\nu$ values, we found that classifiers that trained with uncertain $\Delta\nu$ values were highly robust towards perturbed $\Delta\nu$ values. In addition, high test accuracies in the limit of small $\Delta\nu$ perturbations suggest that the classifiers detect the right set of discriminating features for classification. The accuracy of the classifiers along with the ability to learn discriminating features even for low resolution or noisy data proves that deep learning, as an approach to artificial intelligence, provides a very powerful and robust method to perform asteroseismic classification using power spectra. 
	In the near future, we intend to use our trained classifier to determine the evolutionary state of red giants from all K2 Campaigns and TESS, facilitating progress in stellar astrophysics and enabling a much larger coverage of galactic archaeology studies across the Galaxy than current investigations from the \textit{Kepler} fields.

	\section*{Acknowledgements}
	Funding for this Discovery mission is provided by NASA's Science Mission Directorate. We thank the entire \textit{Kepler} team without whom this investigation would not be possible. D.S. is the recipient of an Australian Research Council Future Fellowship (project number FT1400147). We would also like to thank Timothy Bedding, Daniel Huber, and the asteroseismology group at The University of Sydney for fruitful discussions.
	
	%%%%%%%%%%%%%%%%%%%%%%%%%%%%%%%%%%%%%%%%%%%%%%%%%%
	
	
	
	\bibliographystyle{mnras}
	\bibliography{bibi2} % if your bibtex file is called example.bib


	%%%%%%%%%%%%%%%%%%%%%%%%%%%%%%%%%%%%%%%%%%%%%%%%%%
	
	%%%%%%%%%%%%%%%%% APPENDICES %%%%%%%%%%%%%%%%%%%%%

	%\begin{figure}
	%	\centering
	%	\scalebox{0.4}{\input{/home/z3384751/Downloads/mnras/Diagram2.tex}}	
	%	\caption{Box}
	%\end{figure}
	
	%%%%%%%%%%%%%%%%%%%%%%%%%%%%%%%%%%%%%%%%%%%%%%%%%%

	% Don't change these lines
	\bsp	% typesetting comment
	\label{lastpage}
\end{document}